\documentclass[onecolumn,floatfix,superscriptaddress,a4paper,
               showpacs,showkeys,nofootinbib,preprint]{revtex4}

\usepackage{epsfig}
\usepackage{latexsym}
\usepackage{xspace}
\usepackage{hyperref}
\usepackage[latin2]{inputenc}
\usepackage{indentfirst}
\usepackage{enumerate}

\usepackage{color}

\usepackage{amsmath}
\usepackage{empheq}
\usepackage{amssymb}
\usepackage[english]{babel}
\usepackage{url}

\newcommand{\bs}{\boldsymbol}

\begin{document}

\title{Canonical Ensemble vs. Grand Canonical Ensemble in the
Description of Multicomponent Bosonic Systems}
%
%
\author{D. Anchishkin}
\affiliation{Bogolyubov Institute for Theoretical Physics, 03143 Kyiv, Ukraine}
\affiliation{Taras Shevchenko National University of Kyiv, 03127 Kyiv, Ukraine}
\affiliation{Frankfurt Institute for Advanced Studies, Ruth-Moufang-Strasse 1,
60438 Frankfurt am Main, Germany}
\author{V. Gnatovskyy}
\affiliation{Taras Shevchenko National University of Kyiv, 03127 Kyiv, Ukraine}
\author{D. Zhuravel}
\affiliation{Bogolyubov Institute for Theoretical Physics, 03143 Kyiv, Ukraine}
\author{V. Karpenko}
\affiliation{Taras Shevchenko National University of Kyiv, 03127 Kyiv, Ukraine}
\author{I. Mishustin}
\affiliation{Frankfurt Institute for Advanced Studies, Ruth-Moufang-Strasse 1,
60438 Frankfurt am Main, Germany}
\author{H. Stoecker}
\affiliation{Frankfurt Institute for Advanced Studies, Ruth-Moufang-Strasse 1,
60438 Frankfurt am Main, Germany}
\affiliation{Johann Wolfgang Goethe University, D-60438 Frankfurt am Main, Germany}

\date{\today}



\begin{abstract}

The thermodynamics of a system of interacting bosonic particles and
antiparticles in the presence of the Bose-Einstein condensate is
studied in the framework of the Skyrme-like mean-field model.
It is assumed that the total charge density (isospin density) is conserved
at all temperatures.
Two cases are explicitly considered: zero and nonzero isospin
charge of the system.
A comparative analysis is carried out using Canonical Ensemble and
Grand Canonical Ensemble.
It is shown that the Grand Canonical Ensemble is not suitable for describing
bosonic systems of particles and antiparticles in the presence of a condensate,
but an adequate study can be carried out within the framework of the canonical
ensemble, where the chemical potential is a thermodynamic quantity that depends
on the canonical free variable.
\end{abstract}

\maketitle

\section{Introduction}

The purpose of this letter is to report on the results of studies of the
thermodynamic properties of the bosonic system, in particular, the nature of
phase transitions during Bose-Einstein condensation with conservation of the
isospin (charge) density.
A scalar model of a bosonic system that develops a Bose-Einstein condensate
with conservation of isospin (charge) was first studied in
\cite{haber-1981,kapusta-1981,haber-1982}.
Various aspects of free and interacting systems of relativistic bosons are
discussed further in
Refs.~\cite{bernstein-1991,shiokawa-1999,salasnich-2002,begun-2006,begun-2008,marko-2014}.
To introduce the problem we are going to discuss here, it is appropriate to point
out some features in the description of the condensate phase.
To do this, we first briefly consider Bose condensation in a non-relativistic
ideal bosonic gas within the Grand Canonical Ensemble.
%
At high temperatures where the system is fully in the thermal phase
the density of particles $n$ is
\begin{equation}
n \,=\,  g\int \frac{d^3k}{(2\pi )^3} \, f_{_{\rm BE}}\big(E_k,\mu\big) \,,
\label{eq:nonrel-n}
\end{equation}
where $g$ is the degeneracy factor, $E_k = \bs k^2/2m$
and the Bose-Einstein distribution function $f_{_{\rm BE}}(E,\mu)$ reads
\begin{equation}
f_{_{\rm BE}}(E,\mu) \,=\, \left[ \exp{ \left(\frac{E - \mu}{T}\right) }- 1\right]^{-1} \,.
\label{eq:be-df}
\end{equation}
Here $\mu$ is the chemical potential with some profile, which provides a
conservation of the particle density $n$.
In the phase when the condensate has been developed in the system the total
particle density consists of two contributions, the density of condensed
particles $n_{\rm cond}$ and the density of thermal (kinetic) particles
$n_{\rm th}$:
\begin{equation}
n \,=\, n_{\rm cond}(T) + n_{\rm th}(T)
\qquad {\rm with} \qquad
n_{\rm th}(T) =
g\int \frac{d^3k}{(2\pi )^3} \, f_{_{\rm BE}}\big(E_k,\mu\big)\big|_{\mu = E_{\rm gs}} \,,
\label{eq:nonrel-n-cond}
\end{equation}
where $E_{\rm gs} = 0$ is the energy of the ground state.
In fact, equality $\mu = 0$ is necessary condition for the formation
of condensate.
We consider the homogeneous bosonic system in thermodynamic equilibrium.
In this case, any thermodynamic state of the system in the framework of the
Grand Canonical Ensemble is determined by two canonical variables $(T,\mu)$.
But such a definition of the thermodynamic state cannot be made in the
condensate phase, since we have one free variable $T$ left, and the
chemical potential is fixed by the condition $\mu = 0$.
Thus, to define the thermodynamic state, we need one more variable, which can
be the total particle density $n$.
So, we define the thermodynamic state by two free variables $(T,n)$.
But this means that we describe the bosonic system within the framework of the
Canonical Ensemble, while the Grand Canonical Ensemble is not suitable for
description in the condensed phase.
As Kerson Huang pointed out: "we must re-emphasize that Bose-Einstein
condensation can only occur when the particle number is conserved"
\cite{huang-1987}.
On this way the intersection of the line $n =$~const and the critical
curve $n_{\rm th}(T)$ determines the corresponding critical temperature
$T_{\rm c} = (2\pi/m) (n/g\zeta(3/2))^{2/3}$, which is defined using the
conserved particle density $n$ \cite{landau-v5}.
Hence, the Grand Canonical Ensemble is not very useful for describing the ideal
gas of bosons in the condensate phase.
And what about the use of chemical potential in the thermal phase at high
temperatures?
In fact, one can determine some thermodynamic state of the system by setting
specific values of $(T,\mu)$.
However, to keep the particle density constant, it is necessary to solve
eq.~(\ref{eq:nonrel-n}) with respect to $\mu$ for a given value $n$ and for
temperatures $T > T_{\rm c}$ and get the dependence (profile) $\mu(T,n)$.
But this means that we study the problem again in the Canonical Ensemble,
where the chemical potential is a thermodynamic function depending on the
free variables $(T,n)$ (simply put, using the chemical potential
as a thermodynamic quantity does not mean using the grand canonical
ensemble, see Ref.~\cite{huang-1987}).

A similar situation arises when describing an ideal relativistic bosonic gas with
a constant density of particles $n$.
In this case, the expressions for the density of particles in the high-temperature
thermal phase and in the condensate phase remain the same as before in the
nonrelativistic approach, i.e. eqs.~(\ref{eq:nonrel-n})-(\ref{eq:nonrel-n-cond})
are valid in the relativistic sector.
Only the single-particle ground state becomes $E_{\rm gs} = m$,
since now the dispersion relation is relativistic $E_k = \sqrt{m^2 + \bs k^2}$.
\footnote{Here and below we adopt the system of units $\hbar = c = 1$, $k_{_B} = 1$}
Thus, it follows from the condition of condensate formation that the chemical
potential is equal to the mass of the particle $\mu = m$ in the temperature
interval $0 \le T \le T_{\rm c}$, that is, in the condensate phase.
On the other hand, this value of the chemical potential determines
the maximum density of only thermal particles for a given temperature $T$.

Therefore, it is not possible to determine the thermodynamic state of the
system in the presence of condensate by a pair of variables $(T,\mu)$,
again an additional variable is needed.
It can be the total particle density, then we describe the system by $(T,n)$
variables, i.e., within the Canonical Ensemble.

As one can see, in both cases, with the conserved number of particles, the
Grand Canonical Ensemble cannot describe the bosonic gas in the condensate phase.
This is due to the fact that the chemical potential in this phase is not a
free variable, its value is fixed by the condition of condensate formation,
i.e. $\mu = E_{\rm gs}$.

Next, we are going to prove that the same situation occurs in a system of
interacting relativistic bosons when isospin (charge) is conserved.

\section{The relativistic system of boson particles and antiparticles}
As was stated by Kerson Huang, the real conservation law deals with
the conserved quantity that is the number of particles minus the
number of antiparticles \cite{huang-1987}.
That is why, any study of the Bose-Einstein condensation in the
relativistic Bose gas must take antiparticles into account.
Such systems of boson particles and antiparticles
were first discussed in \cite{haber-1981,kapusta-1981,haber-1982}.
Our consideration of the system of interacting bosonic particles and
antiparticles at finite temperatures is carried out within the framework of
the thermodynamic mean-field model, which was introduced in
\cite{Anch-1992,Anch-1995} and further developed in \cite{Anch-Vovchenko}.
The self-consistent equations for the total particle density $n$ and the
isospin (charge) density $n_I$ in this model are key elements for determining
the thermodynamic state of the system \cite{Anch4,Anch5}.
At high temperatures, when particles ($\pi^{-}$ mesons) and antiparticles
($\pi^+$ mesons) are only in the thermal (kinetic) phase, the corresponding
expressions can be written as follows:
\footnote{We simply conventionally call the particles pi-mesons.
Actually, we consider bosonic particles with a mass of $\pi$-mesons and zero spin. }
\begin{equation}
n =  g\int \frac{d^3k}{(2\pi )^3} \,
\left[ f_{_{\rm BE}}\big(E(k,n),\mu_{I}\big) \,
+\, f_{_{\rm BE}}\big(E(k,n),-\mu_{I}\big) \right] \,,
\label{ntot}
\end{equation}
\begin{equation}
n_{I} =  g\int \frac{d^3k}{(2\pi )^3} \,
\left[ f_{_{\rm BE}}\big(E(k,n),\mu_{I}\big) \,
-\, f_{_{\rm BE}}\big(E(k,n),-\mu_{I}\big) \right] \,.
\label{nI}
\end{equation}
Here the Bose-Einstein distribution functions $f_{_{\rm BE}}\big(E(k,n),\mu_I\big)$,
which is defined in (\ref{eq:be-df}),
contains a term that describes the interaction in the system
$U(n)$: $E(k,n)= \omega_k + U(n),\, \omega_k = \sqrt{m^2 + \bf k^2}$.
It is important to note that in the case of $\mu_I = 0$, the number of particles
and antiparticles in the system are equal to each other, and the net charge in
the system (see Eq.(\ref{nI})) is equal to zero, $n_I = 0$.
This case is discussed in detail in \cite{Anch4}.
There the system of particles and antiparticles was considered using the
Grand Canonical Ensemble.

Below we consider the possibility of Bose-Einstein condensation in a
two-component system, say $\pi^-$ and $\pi^+$ mesons, with a non-zero isospin
(charge) density $n_I = n^{(-)} - n^{(+)}$.
The necessary condition for the formation of condensate is determined by
the maximum possible population of thermal states at a certain temperature,
for example, for particles it is $m + U(n) - \mu_{I} = 0$.
Considering this, the phase structure of the system can be classified according
to three main scenarios:
\\
{\bf (a)} high temperatures - both components, i.e., particles and  antiparticles,
are only in the thermal (kinetic) phase:
$m + U(n) - \mu_{I} > 0$\, and \,  $m + U(n) + \mu_{I} > 0$, respectively.
\\
{\bf (b)} Particles are in the condensate phase
\footnote{``Particles, for example, $\pi^-$-mesons, are in the condensate phase''
is a short name for a mixture phase where one part of the particles is in a
condensate with $\bs k = 0$ but another part of the same particles,
i.e. $\pi^-$-mesons, is in a thermal (kinetic) state.}
and antiparticles are only in the thermal phase:
$m + U(n) - \mu_{I} = 0$\, and \,  $m + U(n) + \mu_{I} > 0$, respectively.
\\
{\bf (c)} Both components, particles and  antiparticles, are in the
condensate phase ($\bs k = 0$):
$m + U(n) - \mu_{I} = 0$\, and \,  $m + U(n) + \mu_{I} = 0$, respectively.
These conditions are equivalent to the system of equations
\begin{empheq}[left=\empheqlbrace]{align}
    & \ \mu_{I}  \,=\, 0 \,,
    \label{muc3}
    \\
    & \ U(n) \,+\, m \,=\, 0 \,.
    \label{muc4}
\end{empheq}

\begin{figure}
\includegraphics[width=0.48\textwidth]{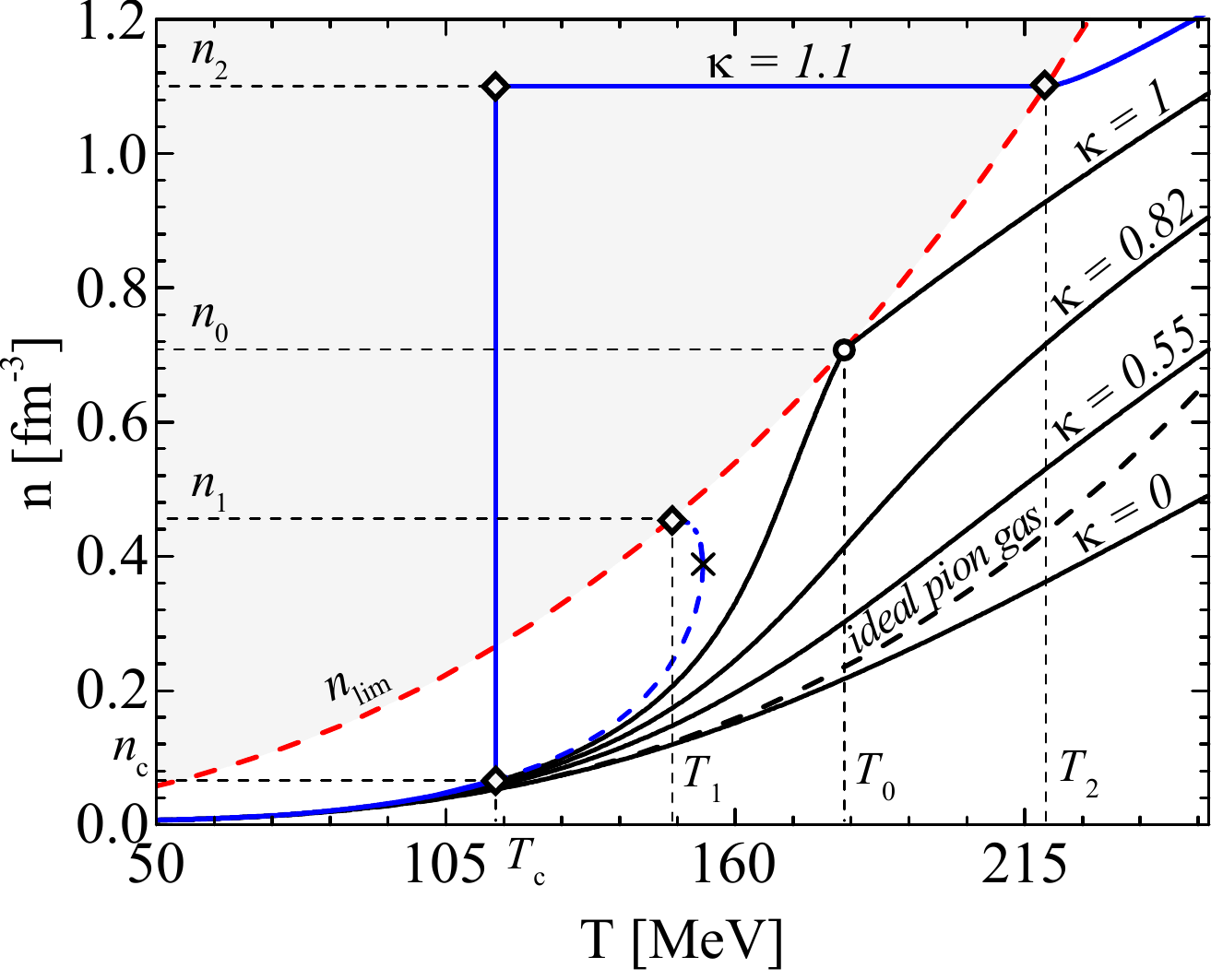}
\hspace{3mm}
\includegraphics[width=0.48\textwidth]{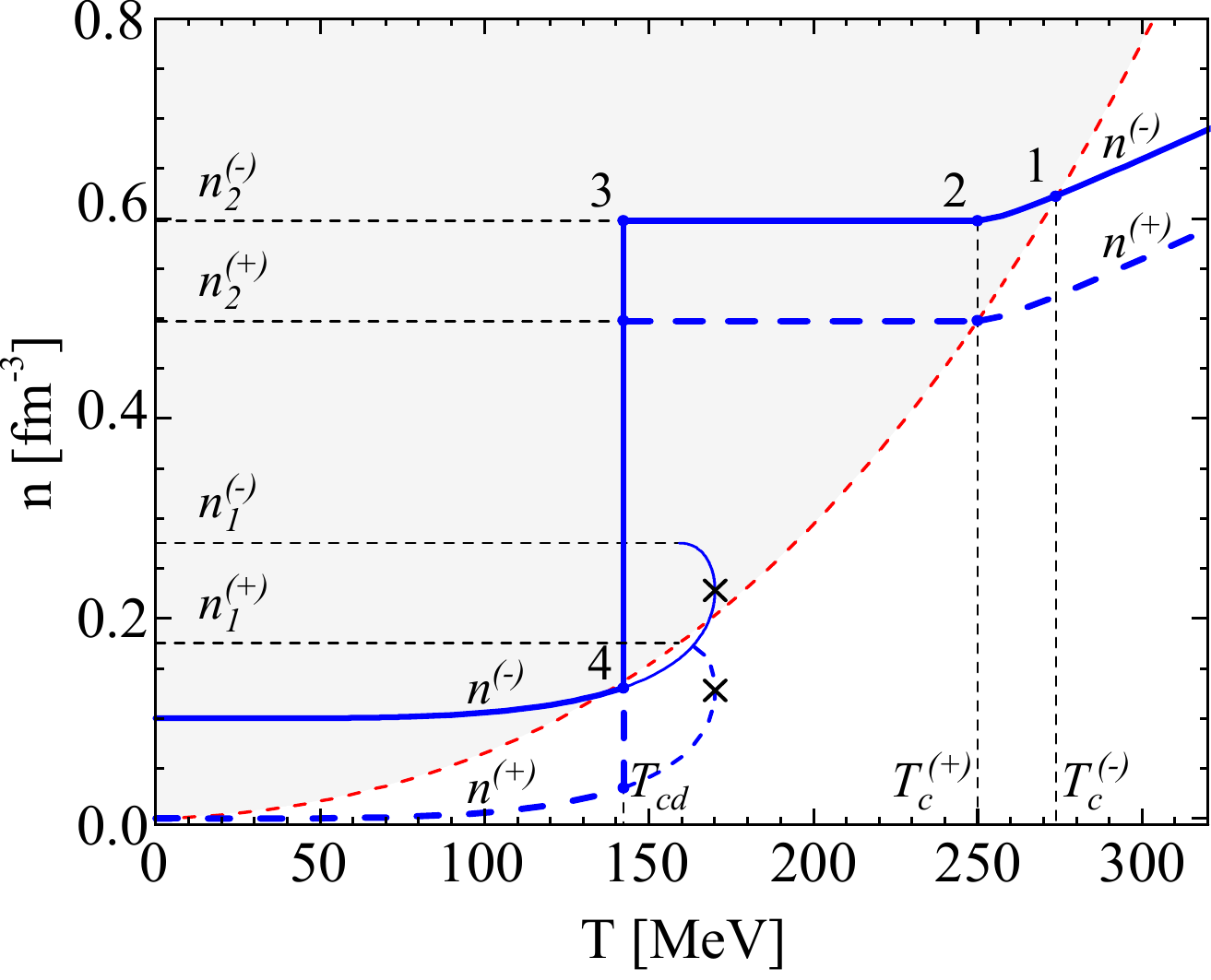}
\caption{\small
{\bf {\it Left panel:}}
Boson system at zero charge density $n_{\rm I} = 0$ (or $\mu_I = 0$).
Particle density vs. temperature at supercritical attraction $\kappa = 1.1$
is shown as a solid blue line consisting of several segments.
The vertical segment (solid blue line) indicates a phase transition of the
first order with the creation of the condensate.
Dependencies of the particle density on the temperature at `weak'' attraction
$\kappa \le 1$ are shown as solid black lines in the thermal phase.
A dashed red line is the critical curve.
{\bf {\it Right panel:}}
Boson system at finite charge density $n_{\rm I} = 0.1$~fm$^{-3}$.
Temperature dependencies of the density of negative $n^{(-)}$ and
positive $n^{(+)}$ particles are shown as a solid blue line consisting of
several segments and a dashed blue line consisting of several segments, respectively.
The vertical segment for both dependencies indicates a phase transition
of the first order with the creation of the condensate.
In the condensate phase $\mu_I = 0$.
A dashed red line is the critical curve $n_{\rm lim}(T)$, see eq.~(\ref{eq:nlim-id}).
 }
\label{fig:id-gas-chem-pot}
\end{figure}

We assume that the interaction between particles is described by the
Skyrme-like mean field, which depends only on the total
particle-number density $n$
\begin{equation}
U(n)\, =\, - \, A\,n\, +\, B\, n^2 \,,
\label{eq:mf1}
\end{equation}
where $A$ and $B$ are the positive model parameters.
For given  mean field (\ref{eq:mf1}) there are two roots of eq.~(\ref{muc4})
with respect to $n$:
\begin{equation}
n_{1,2} \, =\, \sqrt{\frac{m}{B}} \left( \kappa \mp \sqrt{\kappa^2 - 1}\right)\,,
\label{eq:n1-n2-10}
\end{equation}
where
\begin{equation}
\kappa \, \equiv\, \frac{A}{2\,\sqrt{m \,B}} \,.
\label{eq:kappa}
\end{equation}
Then, one can parameterize the attraction coefficient as $A = \kappa A_{\rm c}$
with  $A_{\rm c} = 2\sqrt{m B}$.
The parameter $\kappa$ determines the phase
structure of the system, namely the allowed/forbidden particle density domain.
There are no real roots for the values of the parameter $\kappa < 1$, and we name
this as a ``weak'' attraction.
The critical value  $A_{\rm c}$ is obtained when both roots coincide, i.e.
$\kappa = \kappa_{\rm c} = 1$, then $A = A_{\rm c} = 2\sqrt{m B}$.
And the second interval corresponds to $\kappa > 1$, where eq.~(\ref{muc4})
has two real roots.
We associate this interval with a ``strong'' attractive interaction.
The dependence of the total density of particles $n$ on the parameter $\kappa$
for a system with zero charge $n_I = 0$ is shown in Fig.~\ref{fig:id-gas-chem-pot}
(left panel), which was first obtained in \cite{Anch4} and shown here for comparison
with results calculated for $n_I \ne 0$.

Let us consider several possibilities.
\\
{\bf (a) Both particles and antiparticles are present only in the thermal phase.}
The behavior of the particle-antiparticle boson system in the thermal phase
is determined by a set
of two transcendental equations (\ref{ntot}), (\ref{nI}) (set (a)),
which should be solved selfconsistently with respect to $n$ and $\mu_{I}$
for a given temperature $T$ and $n_I$.
In the present study we consider bosons with spin zero, i.e., the degeneracy
factor $g = 1$ for every boson component.
We would like to point out that in fact we consider the many-particle
system in the Canonical Ensemble, where the independent canonical variables
are $T$ and $n_I$.
In this approach the chemical potential $\mu_{I}$ is a thermodynamic function,
which depends on the canonical variables.

{\bf (b) Particles are in the condensate phase and antiparticles are only in the
thermal phase.}
When the particles are in the condensate phase and antiparticles are
still in the thermal phase, equations (\ref{ntot}), (\ref{nI}) should be
generalized to include condensate component of $\pi^-$ mesons,
$n^{(-)}_{\rm cond}$.
It should be taken into account that the particles (negatively charged component)
can be in a condensed state under the necessary condition
\begin{equation}
 m \,+\, U(n) \,-\, \mu_{I}  \,=\, 0 \,.
\label{eq:condens-cond}
\end{equation}

Let us look on the evolution of the particle-antiparticle system during cooling
it from high temperatures, where the two components are both in the thermal
phase.
When the temperature  decreases from high values, the density of particles
$n^{(-)}(T,n_I)$ first reaches the critical curve at temperature
$T_{\rm c}^{(-)}$, where condition (\ref{eq:condens-cond}) is fulfilled,
see Fig.~\ref{fig:id-gas-chem-pot} right panel.
This condition means that in place of the argument $(E(k,n) - \mu_I)/T$ of
the Bose-Einstein distribution function of charge-dominant component
($\pi^-$ mesons), we put the argument $(\omega_k - m)/T$ and get the curve
$n_{\rm lim}(T)$, defined as
\begin{equation}
n_{\rm lim}(T)  \,=\,  \int \frac{d^3k}{(2\pi)^3}\,
f_{_{\rm BE}}\big(\omega_k,\mu_{I}\big)\Big|_{\mu_{I} = m} \,,
\label{eq:nlim-id}
\end{equation}
which coincide with the critical curve $n_{\rm th}(T)$ for ideal
single-component gas defined in eq.~(\ref{eq:nonrel-n-cond}).
This is indeed a critical curve for $\pi^-$ mesons (see the red dashed curve
in Fig.~\ref{fig:id-gas-chem-pot}, right panel),
since at $\mu_{I} = m$ and temperature $T$ the Bose-Einstein distribution
function reaches its maximum value, which determines the maximum density of
thermal particles at this particular temperature in the system of interacting
bosons.

So, the crossing point of the curves $n^{(-)}(T,n_I)$ and $n_{\rm lim}(T)$
determines the critical temperature $T_{\rm c}^{(-)}$ for high-density component
of the system.
At $T < T_{\rm c}^{(-)}$ the density of thermal $\pi^-$-mesons coincides with
the critical curve (\ref{eq:nlim-id}), i.e. $n^{(-)}_{\rm th} = n_{\rm lim}(T)$.
Thus, the density $n^{(-)}$ of $\pi^-$ mesons
consists of two parts, the density of the condensed $\pi^-$ mesons,
$n^{(-)}_{\rm cond}$,
and the density of thermal $\pi^-$ mesons, $n^{(-)}_{\rm th}$, or
$n^{(-)} = n^{(-)}_{\rm cond}(T) + n_{\rm lim}(T)$.
Therefore, in the temperature interval
$T_{\rm c}^{(+)} \le T \le T_{\rm c}^{(-)}$ we can write the generalization of
the set of eqs.~(\ref{ntot}), (\ref{nI}) (set (b)) as
\begin{eqnarray}
\label{eq:tot-n2a}
n &=&  n^{(-)}_{\rm cond}(T) + n_{\rm lim}(T) \,
+\, \int \frac{d^3k}{(2\pi )^3} \,f_{_{\rm BE}}\big( E(k,n),-\mu_{I} \big)
\Big|_{\mu_{I} = m + U(n)} \,,
\\
n_{I} &=&  n^{(-)}_{\rm cond}(T) + n_{\rm lim}(T)
- \int \frac{d^3k}{(2\pi )^3} \, f_{_{\rm BE}}\big(E(k,n),-\mu_{I}\big)
\Big|_{\mu_{I} = m + U(n)}  \,.
\label{eq:ni2a}
\end{eqnarray}
Solving this system of equations with respect to $n$ and $n^{(-)}_{\rm cond}$
for fixed values of $(T,n_I)$ provides the function $n^{(+)}(T,n_I)$ in the
interval $T_{\rm c}^{(+)} \le T \le T_{\rm c}^{(-)}$, see the blue dashed curve
in Fig.~\ref{fig:id-gas-chem-pot}, right panel.
Obviously, only thermal $\pi^+$ meson contribute to the density $n^{(+)}$.
On the other hand, two fractions of $\pi^-$-mesons contribute to the density
$n^{(-)}$ :
condensed particles with the particle-number density $n^{(-)}_{\rm cond}(T)$,
and thermal particles with the density $n_{\rm lim}(T)$.

{\bf (c) The temperature interval where both particles and antiparticles are in
the condensate phase. }
In this case, additionally to the condition ($\ref{eq:condens-cond}$) for
$\pi^{-}$ mesons, there must be a corresponding condition for $\pi^{+}$ mesons
to ensure that both components of the gas are in the condensate at the same
temperature $T$ and chemical potential $\mu_{I}$.
As we have shown, this requirements lead to equations (\ref{muc3}) and (\ref{muc4}).
With account for these conditions, eqs.~(\ref{eq:tot-n2a}), (\ref{eq:ni2a})
should be modified to include condensate component $n^{(+)}_{\rm cond}$ of
$\pi^{+}$ mesons assuming that the density of thermal $\pi^{+}$ mesons
equals now to $n_{\rm lim}(T)$, i.e. the same way as the density of thermal
$\pi^{-}$ mesons.
So, when both components are in the condensate, set (c) of selfconsistent
equations reads
\begin{eqnarray}
\label{eq:tot-n3}
n &=&  n^{(-)}_{\rm cond}(T) + n^{(+)}_{\rm cond}(T) + 2\,n_{\rm lim}(T) \,,
\\
n_{I} &=&  n^{(-)}_{\rm cond}(T) \,-\, n^{(+)}_{\rm cond}(T) \,.
\label{eq:ni3}
\end{eqnarray}
Indeed, since $\mu_I = 0$, using, for example, the root $n = n_2$ in the
argument of the Bose-Einstein distribution functions in equations
(\ref{eq:tot-n2a}), (\ref{eq:ni2a}) leads to $U(n_2) = - m$, which results in the
density $n_{\rm lim}(T)$, and thus we obtain the system of equations
(\ref{eq:tot-n3}), (\ref{eq:ni3}).
The corresponding particle density of each component $\pi^{-}$ or $\pi^{+}$,
obtained as a result of splitting the total density, for example $n_2$, is equal
to $n^{(-)}_2 = (n_2 + n_{I})/2$ and $n^{(+)}_2 = (n_2 - n_{I})/2$, respectively.

It turns out that solutions to cases (b) and (c) exist in the same
temperature interval.
We are speaking now about the temperature interval $0 < T < T_{\rm c}^{(+)}$,
see Fig.~\ref{fig:id-gas-chem-pot}, right panel.
In addition to self-consistent solutions of equations (b) for these
temperatures, there are two more "condensate" branches of solutions associated
with the roots $n_1$ and $n_2$ of equation $U(n) + m = 0$.
Meanwhile, in the competition between two ``condensate'' branches,
the second branch, created by $n_2$, is preferable,
since the pressure corresponding to these states is higher.
On the other hand,
the competition between solutions (b) and (c) is resolved also in the standard
way according to the Gibbs criterion: the states that correspond to the larger
pressure, are predominant in the thermodynamic realization.
Using this rule we find the temperature $T_{\rm cd}$ from equation
$p_{(b)}(T,n_{I}) = p_{(c)}(T,n_{I})$,  where the pressure $p_{(b)}(T,n_{I})$
corresponds to solutions of the set of equations (b), and $p_{(c)}(T,n_{I})$
to the set of equations (c).
For the temperatures larger than $T_{\rm cd}$ the pressure (c) dominates,
i.e. $p_{(c)}(T,n_{I}) > p_{(b)}(T,n_{I})$.
Therefore, according to the Gibbs criterion, the transition to another branch
of self-consistent solutions gives rise to a phase transition of the first
order at the temperature $T_{\rm cd}$.

Results of the numerical solution of the sets of equations (a), (b) and (c) at
$\kappa = 1.1$ ($B=10mv^{2}_0$, $m=140 \,{\rm MeV}$, $v_0=0.45 \, {\rm fm^3}$)
for the particle-number densities are depicted
in Fig.~\ref{fig:id-gas-chem-pot} in the right panel.
The density $n^{(-)}(T)$ of $\pi^-$ mesons depicted as a solid blue
curve that consists of several horizontal segments and one vertical segment,
which represents a first-order phase transition.
The density $n^{(+)}(T)$ of $\pi^+$ mesons is depicted as a dashed blue line,
which also consists of several horizontal segments and one vertical segment,
which also represents phase transition of the first order.
Metastable and forbidden states belonging to ``thermal'' solutions are
shown in Fig.~\ref{fig:id-gas-chem-pot} in both panels as dashed and dotted
segments, resembling extensions of thermal branches.
The right panel of the figure shows that the system is actually described
with a conserved isospin (charge) density,
i.e., for each temperature point on the graph, the
condition $n^{(-)}(T) - n^{(+)}(T) = 0.1$~${\rm fm^{-3}}$ is fulfilled.
The conservation of charge leads to the splitting of the roots
(\ref{eq:n1-n2-10}) shown in the left panel.
It looks like:
$n_1 \to n^{(-)}_{1}, \ n^{(+)}_{1}$ and
$n_2 \to n^{(-)}_{2}, \ n^{(+)}_{2}$, where the difference is constant,
for example $n^{(-)}_2 - n^{(+)}_2 = n_I$.

In both panels of Fig.~\ref{fig:id-gas-chem-pot}, in the condensate phase,
both systems are represented by {\it zero chemical potential} regardless of
whether the particle-antiparticle system described in the right panel has a
finite charge density, i.e., $n_I = 0.1$~$\rm fm^{-3}$, while the
particle-antiparticle system described in the left panel is characterized by
zero charge density, i.e., $n_I = 0$.
So if one intends to study both systems, one system with a finite charge density
and another with zero charge density within the Grand Canonical Ensemble, then
the canonical variables should be $(T,\, \mu_I = 0 )$ when describing the
condensate phase in both systems.

It seems that we come to some contradiction, because in the text books it is
written that the chemical potential should reflect the conservation of charge
or conservation of the particle number.
This contradiction is resolved by realizing
that the Grand Canonical Ensemble with the canonical variables $(T,\, \mu)$ is
suited for the description of the thermal phase or for a description of the
particles, which are in the kinetic states, but not in the condensate states.
Indeed, it can be seen in Fig.~\ref{fig:id-gas-chem-pot} on the right panel in
the temperature interval that corresponds to the condensate phase, that is,
between points 2 and 3 on the graph, that for each temperature from this
interval the density of thermal $\pi^-$ mesons is equal to the density
of thermal $\pi^+$ mesons, both equal to $n_{\rm lim}(T)$ defined in
(\ref{eq:nlim-id}).
In other words, these two densities of thermal mesons, which are
characterized by $\mu_I = 0$, are equal to the density on the critical
curve $n_{\rm lim}$, which is depicted as the dashed red line
in Fig.~\ref{fig:id-gas-chem-pot} (right panel).
That is why the charge density, which is determined only by thermal
particles, is zero, i.e., $n_I^{\rm (therm)} = n_{\rm lim}(T) - n_{\rm lim}(T) = 0$.
Therefore, the chemical potential, which corresponds to the charge of these
thermal particles, is also zero.
We can conclude that the chemical potential $\mu_I$ is a useful quantity only
for describing thermal or kinetic particles.
This automatically leads to the fact that the Grand Canonical Ensemble, where
the chemical potential $\mu_I$ is a free variable, can be adequate to describe
the bosonic system only in the thermal phase.
In the condensate phase, an adequate tool for describing this phase is the
Canonical Ensemble, where the chemical potential $\mu_I(T,n_I)$ is a
thermodynamic quantity that depends on free variables.
It should be noted that this statement fully corresponds to our conclusion
about the description of the ideal bosonic gas in the condensate phase obtained
in the Introduction.

{\bf Other examples} \\
One can argue that the mean field depends on the isospin density.
Indeed, consider the thermodynamic mean-field model, where the mean field
depends on the total particle density $n$ and also depends on the isospin
density  $n_I$ (this possibility was discussed in \cite{Anch5}).
As shown in Ref.~\cite{Anch-1995}, since $n$ and $n_I$ are independent
thermodynamic variables, the form of this mean field is as follows:
$U^{(\mp)}(n,n_{I}) = U(n) \mp U_{I}(n_{I})$, where $U_I(n_I)$
is an odd function, for example, $U_I(n_I) \propto n_I$, and the field
$U^{(-)}$ acts on $\pi^-$ mesons, while $U^{(+)}$ acts on $\pi^+$ mesons.
Then, if $\pi^-$ and $\pi^+$ mesons are in the condensate phase, two
necessary conditions must be fulfilled:
$m + U(n) - U_{I}(n_{I}) - \mu_I = 0$ and $m + U(n) + U_{I}(n_{I}) + \mu_I = 0$.
From here we get the equivalent equations: $m + U(n) = 0$ and
$\mu_I = - U_{I}(n_{I})$.
Therefore, the chemical potential is fixed by the condition of condensate
formation, and is determined by the isospin density, which remains constant.
Hence, when the mean interaction in the system depends on the isospin (charge)
density, we again conclude that $\mu_I$ cannot be a free variable in the
presence of a condensate, and hence the Grand Canonical Ensemble is not
applicable in the condensate phase.

When describing the same multi-boson system at a finite charge density in the
field-theoretic approach \cite{mishustin-anchishkin-2019}, we encounter
exactly the same paradox.
Indeed, for developing of the condensate by both particles and
antiparticles, two conditions must be met: $M - \mu_I = 0$ and $M + \mu_I = 0$,
where $M$ is the effective mass of quasiparticles.
By complete analogy with case (c) discussed above, these conditions lead to
two equations: $M = 0$ and $\mu_I = 0$.
So, it turns out that the system with a finite charge density $n_I \ne 0$ is
characterized by zero value of the chemical potential.
On the other hand, we see that in the presence of condensate, the density of
thermal particles is the same in the negatively and positively charged
components of the system, i.e., $n^{(-)}_{\rm th}(T) = n^{(+)}_{\rm th}(T)$.
Hence, the problem can be resolved by accepting that the chemical potential is
responsible only for thermal (kinetic) particles.

The picture obtained becomes even more striking when we study the
conservation of charge
in a relativistic ideal boson gas of particles and antiparticles at $n_I \ne 0$.
Indeed, if we assume that particles and antiparticles are simultaneously in the
condensate phase, then two conditions must be satisfied simultaneously:
$m - \mu_I = 0$ and $m + \mu_I = 0$,
where $\mu_I$ is the isospin chemical potential, which corresponds to $n_I$.
This leads to two equations: $m = 0$ and $\mu_I = 0$.
As we can see, the first equation is not possible or unphysical.
That is, only one condition can be fulfilled, for example $m - \mu_I = 0$.
So, we can formulate the theorem:
{\it In a relativistic bosonic ideal gas of particles and antiparticles with
a conserved isospin (charge) $n_I \ne 0$, only one component of the system
can form a condensate phase.
In the case of zero isospin $n_I = 0$, the system of particles and antiparticles
does not form a condensate.}
The dashed black line in Fig.~\ref{fig:id-gas-chem-pot} (left panel) can be
seen as an illustration of the second statement.

Almost the same features in the behavior of the ideal relativistic
particle-antiparticle bosonic gas were noticed by H.E.~Haber and H.A.~Weldon
in 1981 in Ref.~\cite{haber-1981}:
below $T_{\rm c}$ the chemical potential determines only the charge density of
the excited states $\rho - \rho_0$, where $\rho_0$ is the charge density of the
ground state.
Which actually means that the boson system in the condensate phase cannot be
adequately described by a Grand Canonical Ensemble, but can be described by
a Canonical Ensemble, where $\mu(T,n_I)$ (according to our notation:
$\rho = n_I$, \, $\rho_0 = n_{\rm cond}$).
Also, in \cite{haber-1982} they write that "since $n_I$ is a physical
quantity and $\mu_I$ is a derived quantity", equation (\ref{nI}) is in fact
an implicit formula for $\mu_I$ as a function of $n_I$ and $T$.
This means that the Canonical Ensemble is used.

In addition, in the thermodynamic mean-field model considered above,
we have similar conditions for the formation of condensate by both
components: $m + U(n) - \mu_I = 0$ and $m + U(n) + \mu_I = 0$.
When the effective interaction in the system vanishes, i.e., $U(n) \to
0$, these conditions are the reason for the same statement that only
one component of the bosonic particle-antiparticle system can
develop a condensate.
And as we saw above, both components can be in
a condensed state only with the help of the attractive mean field,
when its value is equal to the mass of the particle, i.e. $m = |U(n)|$.

{\bf Conclusion} \\
We have shown how the relativistic interacting system of Bose particles
and antiparticles can be described at zero and finite isospin (charge)
density.
In both cases
the meson system develops a first order phase transition for sufficiently
strong attractive interactions via forming a Bose condensate with releasing
the latent heat.
The model predicts that the condensed phase is characterized by a constant
density of particles.

We have demonstrated that the Grand Canonical Ensemble is {\it not suitable}
for describing a multi-component bosonic system in the presence of condensate
phase.
In particular, it cannot describe the condensate state in the system of
particles and antiparticles.
The reason is that the chemical potential is not a free parameter
in the condensate phase, its values are determined by the
necessary condition for condensate formation.
As we have shown, these statements are valid in interacting bosonic systems
as well as in an ideal bosonic gas.

Details of the calculations presented in this letter can be found
in \cite{universe-2023}.

\section*{Acknowledgements}
D.A. is very grateful to J.~Steinheimer and O.~Philipsen for useful discussions
and comments and greatly appreciates the warm hospitality and support provided
by FIAS administration and the scientific community.
The work of D.Zh. and D.A. was supported by the Simons Foundation and by
the Program
"The structure and dynamics of statistical and quantum-field systems" of
the Department of Physics and Astronomy of the NAS of Ukraine.
I.M. thanks FIAS for support and hospitality.
H.St. thanks for support from the J. M. Eisenberg Professor
Laureatus of the Fachbereich Physik.


\end{document}